\documentclass[preprint,preprintnumbers]{revtex4}

\usepackage{graphicx}

\usepackage{amssymb}

\usepackage{mathrsfs}

\begin{document}

\title{Interference and  transport properties of conductions  electrons   on the surface of  a topological insulator}

\author{D. Schmeltzer }

\affiliation{Physics Department, City College of the City University of New York 
New York, New York 10031, USA}

\begin{abstract}

The   surface conductivity  for    conduction electrons  with a fixed chirality in  a topological insulator with impurities scattering   is considered.
The surface excitations are described  by the Weyl Hamiltonian.  For a finite  chemical potential  one  projects out the hole band and one obtains   a single  electronic band with a fixed   chirality.  One obtains a model of spinless electrons which experience a half vortex when they   return to the origin.
As a result the conductivity is equivalent to a spinless problem with correlated noise which gives rise to anti-localization. 
We compute  conductivity as a function of  frequency   and   compare our  results with the  $Raman$ shift measurement   for  $Bi_{2}Se_{3}$.


\end{abstract}

\maketitle
\textbf{I. Introduction}

 Topological insulators ($TI$)  are time reversal invariant systems  which  obey Kramers  theorem   \cite{Volkov,Kane,Zhang,David}. 
Toplogical insulators  are characterized    by the  surface  excitations   and are described by the  $Weyl$ Hamiltonian.  $Bi_{2}Se_{3}$  is such a material  which is  described by the  $Weyl$  equation with a single Dirac cone  which lies below the chemical potential $\mu$ and the bulk gap.   Due to time reversal symmetry  the backscattering is suppressed  in agreement  with  scanning tunneling  microscopy (STM) experiment and theory \cite{Balatsky}.  Conductivity results   are less conclusive, due to the presence of the  $3D$  bulk gap  \cite{Culcer} or  the insulating gap   observed in   thin layers of $TI$.

The charge  current  for the  Weyl  equation  is identified with the  spin half   operator.  As a result the charge current and the spin current are related to each other \cite{Raghu}.
In the presence of impurities,  elastic scattering conserves energy but not quasi-momentum  giving rise to a finite  conductivity.

Transport properties have been calculated  building on the relation with the spin-orbit model,  which belongs to the symplectic  ensemble \cite{Hikami} and therefore quantum interference gives  rise to anti-localization. These results have been confirmed by \cite{Ando}  and  \cite{Hankiewicz,Stern,Shen,Garate}.

The purpose of this work  is to compute  the surface conductivity for the   Weyl Hamiltonian  with   a finite chemical potential $\mu$.  The surface Hamiltonian  for a single Dirac cone and a finite chemical potential has been obtained in ref. \cite{Raghu} and can be described by the Weyl Hamiltonian \cite{Maggiore} with a Fermi velocity $v\approx 5\times 10^{6}$  $\frac{m}{sec}$ and  a chemical potential  $\mu\approx 0.1$ eV. 
The Weyl Hamiltonian is characterized by  opposite chirality for electrons and holes.
For a large chemical potential we can replace the Weyl Hamiltonian by a single electronic band with a fixed  chirality.
The model resembles a spinless model in two dimensions, therefore the presence of elastic scattering might give rise to localization.   The backscattering potential  for a single scattering  is zero  but multi-scattering  are allowed. Therefore we might expect localization . Consequently,  a particle can backscatter  after  two  scatterings allowing a decrease in the conductivity.
The difference between  the one band spinless electrons  and our case   arising from  the fact that the conduction electrons are obtained after a projection of the chiral spinors of the Weyl  spin half fermions. The effect of the projection modifies the random scattering matrix elements  but preserve the property of spin half spinor introducing correlations which are determined by the momentum difference between the incoming and outgoing electrons. 
In order to determine the quantum corrections   we compute  the  quantum return probability  $P_{qm}$ given by   the   interference between  a closed path and the time reversed path. Due to the projection which preserve the spin half vortex at  $\vec{k}=0$  the quantum return probability  $P_{qm}$ vanishes.
Using these results  we compute the conductivity using the method employed for spinless electrons. The  impurity scattering   affects  the  charge  and  gives rise to the maximal crossed corrections \cite{Rammer}. We find that the charge is enhanced giving rise to an enhanced  transport life time.  Due to the angle dependent  scattering  the maximal crossed diagrams  change sign and give rise  to anti-localization. This effect  result is similar  with the results obtained by \cite{Ando} in graphene  with a zero chemical potential.

The plan of this paper is as follows. 
In Sec. $II$ we introduce the Weyl model.  We construct the quantized particle and anti-particle bands imposing the time reversal symmetry. Due to the singularity at $\vec{k}=0$ the unperturbed spinor must  be obey the the  Pfaffian  properties \cite{Kane}.  In Sec. $III$ we consider the model for a single conduction band .In Sec. $IV$   we compute the quantum return probability for the one band model. In Sec. $V$ we construct the Green's functions \cite{Abrikosov} for  the conduction electrons.  We  compute the $ladder$ and $maximal$ $crossed$ diagram . In Sec. $VI$ we compute the static and frequency  dependent  conductivity. In Sec. $VII$ we discuss our results and propose that a good confirmation of the surface conductivity can be obtained from Raman scattering \cite{Conder}.In Sec. $VIII$ we present our conclusions.

\vspace{0.2 in}

\textbf{II. The Weyl model }

\vspace{0.2 in}

The surface state Hamiltonian  for a topological insulator of the  $Bi_{2}Se_{3}$ family materials  is given by the Weyl model \cite{Raghu}.
The presence of a  random potential and an  electromagnetic field  modify the Weyl model  in the following way:
\begin{eqnarray}
&&S=\int\,dt \left[\int\,d^2 r\Psi^{\dagger}(\vec{r},t)(i\partial_{t} -eA_{0}(\vec{r},t))\Psi(\vec{r},t)-H\right], \nonumber\\ 
&&H=\hbar v  \int\,d^2 r \Psi^{\dagger}(\vec{r}) \left[\sigma^{1}(-i\partial_{2}-\frac{e}{\hbar}A_{2}(\vec{r},t))-\sigma^{2}(-i\partial_{1}-\frac{e}{\hbar}A_{1}(\vec{r},t))+ U_{sc}(\vec{r})\right]\Psi(\vec{r}) , \nonumber\\&&
\end{eqnarray}
where $\vec{A}(t)=\frac{\vec{E}(\Omega)}{i \Omega}e^{-i\Omega t}$
is the external  vector potential, $A_{0}(\vec{r},t)$ is the external scalar potential and  $U_{sc}(\vec{r})$ is the random potential controlled by the   white noise correlation function  
$\langle\langle U_{sc}(\vec{k}) U_{sc}(\vec{k}')\rangle\rangle=D_{sc}\delta^{2}[\vec{k}+\vec{k}']$. 
The unperturbed  Weyl    Hamiltonian in the first quantized form is given by  $h[\vec{k}]=-\sigma^{2}k_{1}+\sigma^{1}k_{2}$.   Here $h[\vec{k}]$ 
is time reversal invariant  and obeys the transformation  relation $\vartheta^{-1}h[\vec{k}]\vartheta =h[-\vec{k}]$,  where   $\vartheta=i\sigma^{2}K$ is the time reversal operator and $K$ is the conjugation operator. The eigenstates form a Kramers degenerate pair with the property $\vartheta^2=-1$.
At the point $\vec{k}=0$  the eigenvectors  $|u^{(+)}(\vec{k})\rangle$, $|u^{(-)}(\vec{k})\rangle$  are degenerate and we need to choose a representation where the eigenvectors are orthogonal to each other. 
In order to regularize the problem we will replace the Hamiltonian $h[\vec{k}]=-\sigma^{2}k_{1}+\sigma^{1}k_{2}$ by:
\begin{equation}
h[\vec{k}]=-\sigma^{2}k_{1}+\sigma^{1}k_{2}+ \sigma^{3}\eta M(\vec{k}), \hspace{0.1 in} \eta\rightarrow 0 . 
\label{ham}
\end{equation}
We introduce the function $\theta[M(\vec{k})]$ which is one for $M(\vec{k})\geq 0$ and zero otherwise. We solve  the eigenvalue   equation in the presence of the mass term $M(\vec{k})$ . For the region   $\theta[M(\vec{k})]=1$  find that that  the eigenvector $|u_{1}(\vec{k})\rangle$ corresponds to the positive eigenvalues   and eigenvector $|u_{2}(\vec{k})\rangle$ to the negative eigenvalues .  For the region  $\theta[-M(\vec{k})]=1$ the eigenvector  $|u_{2}(\vec{k})\rangle$ corresponds to  positive eigenvalues  and  $|u_{1}(\vec{k})\rangle$  correspond to the negative eigenvalues. Using a  mass term   which preserve time reversal symmetry  \textbf{$M(\vec{k})=-M(-\vec{k})$}, we guarantee that  the eigenvectors at the   point $\vec{k}=0$  are orthogonal to each other and obey Krammers theorem. We will consider the limit $\eta\rightarrow 0$ and  neglect the energy corrections but we will consider  the topological effects  caused by the mass term  $M(\vec{k})$ which preserve the time reversal symmetry  $\vartheta^2=-1$.  We note that for the  $wrapping$  case,  the mass $M(\vec{k})=(k_{1}+ik_{2})^3+(k_{1}-ik_{2})^3=2k_{1}(k^2_{1}-3k^2_{2})$ provide  a natural regularization  for the eigenvectors at $\vec{k}=0$ .
The eigenvectors  $|u_{1}(\vec{k})\rangle$ and  $|u_{2}(\vec{k})\rangle$ are given in terms of the multivalued phase   phase  $\chi(\vec{k})$, $\tan(\chi(\vec{k})=\frac{k_{2}}{k_{x}}$  :
\begin{equation}
|u_{1}(\vec{k})\rangle=\frac{1}{2}\Big[|\vec{k}\rangle\otimes[1,-ie^{i\chi(\vec{k})}]^{T}\Big], \hspace{0.2 in}
|u_{2}(\vec{k})\rangle=\frac{1}{2}\Big[|\vec{k}\rangle\otimes[ie^{-i\chi(\vec{k})},1]^{T}\Big] , 
\label{state}
\end{equation}
The phase $\chi(\vec{k})$ obey the property $\chi(-\vec{k})=\pi+\chi(\vec{k})$.
$|u_{1}(\vec{k})\rangle$ and  $|u_{2}(\vec{k})\rangle$  form a    Kramers  pair.  Due to the  time reversal symmetry   the two eigenvectors   are related through the Pfaffian matrix, $W_{1,2}(\vec{k})=\langle u_{1}(\vec{k})|\vartheta|u_{2}(\vec{k})\rangle$ with $|u_{1}(-\vec{k})\rangle= W^{*}_{1,2}(\vec{k})\vartheta  u_{2}(\vec{k})\rangle$ and   $|u_{2}(-\vec{k})\rangle=W^{*}_{2,1}(\vec{k})\vartheta|u_{1}(\vec{k})\rangle$.
The spinors $|u_{1}(\vec{k})\rangle$ and $|u_{2}(\vec{k})\rangle$   will  be used to construct the eigenvector  $|u^{(+)}(\vec{k})\rangle$ for particles  and $|u^{(-)}(\vec{k})\rangle$ for antiparticles for the entire Brillouin zone.
\begin{eqnarray}
&&|u^{(-)}(\vec{k})\rangle=\theta[M(\vec{k})]|u_{2}(\vec{k})\rangle+ \theta[M(-\vec{k})]|u_{1}(\vec{k}\rangle ,\nonumber\\&&
|u^{(+)}(\vec{k})\rangle=\theta[M(\vec{k})]|u_{1}(\vec{k})\rangle+ \theta[M(-\vec{k})]|u_{2}(\vec{k})\rangle . \nonumber\\&&
\end{eqnarray}
The eigenvector  $|u^{(+)}(\vec{k})\rangle$ corresponds to the  eigenvalue $E_{p}=E_{+}=\hbar v |\vec{k}|$  (for particles)  and    $|u^{(-)}(\vec{k})\rangle$ corresponds to  the eigenvalue  $E_{a}=E_{-}=-\hbar v |\vec{k}|$ (for  anti-particle).  The eigenspinors  chage  the sign under a full rotation of  $2\pi$. In two dimensions a full rotation of $2\pi$  is equivalent to two consecutive inversions. We introduce the   inversion operator $P$, $P|u^{(+)}(\vec{k})\rangle= |u^{(+)}(-\vec{k})\rangle$, $P|u^{(-)}(\vec{k})\rangle= |u^{(-)}(-\vec{k})\rangle$ and find  that $P^2=-1$:
\begin{eqnarray}
&&P^2e^{\frac{\pm i}{2}\chi(\vec{k})} =P e^{\frac{\pm i}{2}\chi(-\vec{k})}=P e^{\frac{\pm i}{2}(\chi(\vec{k})+\pi)}= e^{\frac{\pm i}{2}(\chi(-\vec{k})+\pi)}=e^{\frac{\pm i}{2}(\chi(\vec{k})+\pi+\pi)}= e^{i\pi}e^{\frac{\pm i}{2}\chi(\vec{k})}=-e^{\frac{\pm i}{2}\chi(\vec{k})}\nonumber\\&&
P^2|u^{(+)}(\vec{k})\rangle=-|u^{(+)}(\vec{k})\rangle ;  P^2|u^{(-)}(\vec{k})\rangle=-|u^{(-)}(\vec{k})\rangle \nonumber\\&&
\end{eqnarray}
This shows that the  two eigenspinors  $|u^{(\pm)}(\vec{k})\rangle $ behave like  a spin half fermion under rotations. 

We will  include the  Fermi velocity $v$ to describe the physical energy spectrum. In order to incorporate the velocity in the Hamiltonian we  multiply the spinor operator by $\sqrt{v}$.
The spinor operator $\Psi(\vec{r})$  is decomposed into the eigenmodes of the unperturbed  Weyl    Hamiltonian. 
\begin{equation}
\Psi(\vec{r})=\sqrt{v}\sum_{\vec{k}}e^{i\vec{k}\cdot \vec{r}}[C(\vec{k})u^{(+)}(\vec{k})+ B^{\dagger}(-\vec{k})u^{(-)}(-\vec{k})]\equiv \sqrt{v} \sum_{\vec{k}}e^{i\vec{k}\cdot \vec{r}}\Psi(\vec{k}) . 
\label{gr}
\end{equation}
We choose that  particle operators $C(\vec{k})$, $C^{\dagger}(\vec{k})$   and the anti-particle operators  $B(\vec{k})$, $B^{\dagger}(\vec{k})$ should  obey anti-commutation relations:
\begin{equation}
[C(\vec{k}),C^{\dagger}(\vec{k'})]_{+}=[B(\vec{k}),B^{\dagger}(\vec{k'})]_{+}  =\delta^{(2)}[\vec{k}-\vec{k'}],\hspace{0.1 in} [C(\vec{k}),C(\vec{k'})]_{+}=[B(\vec{k}),B(\vec{k'})]_{+}=0 . 
\label{opq}
\end{equation}
The spinor projection $\Psi_{\sigma}(\vec{r})\equiv \langle\Psi(\vec{r})| \sigma \rangle$ obeys the  modified  anti-commutation relations:
\begin{equation}.
[\Psi_{\sigma}(\vec{r}),\Psi^{\dagger}_{\sigma'}(\vec{r'})]_{+}=v\delta_{\sigma,\sigma'}\delta^{(2)}(\vec{r},\vec{r'}),\hspace{0.1 in} [\Psi_{\sigma}(\vec{r}),\Psi_{\sigma'}(\vec{r'})]_{+}=0 , 
\label{anti}
\end{equation}
For a a chemical potential  $\mu= \hbar v k_{F} >0$  ($k_{F}$  is the Fermi momentum) we introduce the notation  $|\mu>$ for the  ground state.  The particle operator  $C(\vec{k})$ and the  anti-particle operator $B(\vec{k})$ annihilate the ground state, $C(\vec{k})|\mu\rangle= 0$  for $ E(\vec{k})>\mu>0$.
We will consider a situation  where the  chemical potential $\mu$   measured from the Dirac point  has the typical value of $0.01$  eV with a Fermi velocity $v$ of $5\times 10^{5}$ $\frac{m}{sec}$. Using the eigenspinors  given in eq. $(4)$  we find that  the Hamiltonian  in the present of the scattering potential  $U_{sc}(\vec{k}-\vec{p})$ is equivalent to two coupled bands.
\begin{eqnarray}
&&H_{0}=\sum_{\vec{k}}[(\hbar v|\vec{k}|-\mu)C^{\dagger}(\vec{k})C(\vec{k})+ (\hbar v|\vec{k}|+\mu)B^{\dagger}(-\vec{k})B(-\vec{k})] , \nonumber\\&&
H_{D}=\sum_{\vec{k}}\sum_{\vec{p}}v U_{sc}(\vec{k}-\vec{p})\Big[ V^{(p,p)}(\vec{k}, \vec{p})C^{\dagger}(\vec{k})C(\vec{p})+V^{(a,a)}(-\vec{k}, -\vec{p})B(-\vec{k})B^{\dagger}(-\vec{p})\nonumber\\&& +
V^{(p,a)}(\vec{k}, -\vec{p}) C^{\dagger}(\vec{k})B^{\dagger}(-\vec{p})+V^{(a,p)}(-\vec{k}, \vec{p})B(-\vec{k})C(\vec{p})\Big] , \nonumber\\&&
\end{eqnarray} 
where $H_{0}$ is the unperturbed Hamiltonian and $H_{D}$ represent the effect of the impurity scattering.
The spinors structure gives rise to  vertex functions for the coupling to the random potential in momentum space. The vertex functions are given in terms of the particle and anti-particle matrix elements ($p$ stands for particles and  $a$ stands for antiparticles): 
\begin{eqnarray}
&& V^{(p,p)}(\vec{k}, \vec{p})=\langle u^{(+)}(\vec{k})|u^{(+)}(\vec{p})\rangle, ~~  V^{(a,a)}(-\vec{k}, -\vec{p})=\langle u^{(-)}(-\vec{k}))|u^{(-)}(-\vec{p})\rangle, \nonumber\\&&
 V^{(p,a)}(\vec{k}, -\vec{p})=\langle u^{(+)}(\vec{k})|u^{(-)}(-\vec{p})\rangle, ~~  V^{(a,p)}(-\vec{k}, \vec{p})=\langle u^{(-)}(-\vec{k})|u^{(+)}(\vec{p})\rangle . \nonumber\\&&
\end{eqnarray}
From Eq. $(1)$  we obtain the external coupling of the electrons to the electromagnetic field $\vec{A}(\vec{r},t)$: 
\begin{equation}
H^{ext}(t)=\int\,d^{2}r [\hat{J}_{1}(\vec{r},t)A_{1}(\vec{r},t)+\hat{J}_{2}(\vec{r},t)A_{2}(\vec{r},t)] . 
\label{eq}
\end{equation}
From  Eq. ($1$)  we also find that the currents are given by: 
$\hat{J}_{1}(\vec{r},t)\equiv (-e)v \Psi^{\dagger}(\vec{r})\sigma^{2}\Psi(\vec{r})$,  $\hat{J}_{2}(\vec{r},t) \equiv(-e)v\Psi^{\dagger}(\vec{r})(-\sigma^{1})\Psi(\vec{r}).$
 
Using the linear response theory \cite{Doniach} for the current-current correlation function in the presence of  an external vector potential  $A_{1}(t)=\frac{ E_{1}}{i\Omega}e^{-i\Omega t}$  we compute the   interband  absorbtion. We find that  the optical conductivity in the limit of weak disorder   and  frequencies  $\Omega > 2\mu $ is given by  the universal relation   $\sigma(\Omega > 2\mu)=\frac{e^2\pi}{8h}$. 

For  a  finite chemical potential $\mu$ we  integrate the anti-particle band $ H^{(a)}$: 
\begin{eqnarray}
&& H^{(a)}= \sum_{\vec{k}}(\hbar v|\vec{k}|+2 \mu)B^{\dagger}(-\vec{k})B(-\vec{k})+
\sum_{\vec{k}}\sum_{\vec{p}}v U_{sc}(\vec{k}-\vec{p})\Big[V^{(a,a)}(-\vec{k}, -\vec{p})B(-\vec{k})B^{\dagger}(-\vec{p})  \nonumber\\&&+
V^{(p,a)}(\vec{k},-\vec{p}) C^{\dagger}(\vec{k})B^{\dagger}(-\vec{p})+V^{(a,p)}(-\vec{k}, \vec{p})B(-\vec{k})C(\vec{p})\Big] . \nonumber\\&& 
\end{eqnarray}
 We perform the integration in the limit where white noise  fluctuations are negligible in comparison with the chemical  potential $\sqrt{D_{sc}}< \mu$. Under this condition the anti-particles can be integrated out. As a result  the particle Hamiltonian is modified by the induced term $\delta H^{ind.}$ which can be understood as shift of the chemical potential.
\begin{eqnarray}
&&\delta H^{ind.}=-\int\frac{d^2k}{2\pi}\int\frac{d^2q}{2\pi}\int\frac{d\omega}{2\pi} C^{\dagger}(\vec{k},\omega)\nonumber\\&&\left[\int\frac{d^2p}{2\pi}V^{(p,a)}(\vec{k},\vec{p})\frac{ v U_{sc}(\vec{k}-\vec{p})U_{sc}(\vec{p}-\vec{k}+\vec{q})v}{[(E(\vec{k})+\Sigma(\vec{k},\omega;p))-(E(\vec{p})+ \Sigma(\vec{p},\omega;a)) -2\mu] }V^{(a,p)}(-\vec{p},\vec{k}-\vec{q})\right]C(\vec{k}-\vec{q},\omega). \nonumber\\&&
\end{eqnarray}
Here $\Sigma(\vec{k},\omega;a)$ is the anti-particle self-energy induced by the disorder. When the chemical potential $\mu$ obeys the condition $|(E(\vec{k})+\Sigma(\vec{k},\omega;a))-(E(\vec{p})+ \Sigma(\vec{p},\omega;a)) -2\mu|\leq 2\mu$ we can perform the momentum $p$ integration which allows  the replacement of $U_{sc}(\vec{k}-\vec{p})U_{sc}(\vec{p}-\vec{k}+\vec{q})$ with $\langle\langle U_{sc}(\vec{k}-\vec{p})U_{sc}(\vec{p}-\vec{k}+\vec{q})\rangle\rangle=D_{sc}\delta^{2}[\vec{q}]$. Consequently, the effective one-band Hamiltonian  will have a  new chemical potential $ \mu\rightarrow \mu-\frac{D_{sc}v^{2}}{2\mu}$. Therefore for chemical potential $\mu$ which satisfies $\mu>\frac{D_{sc}}{2}$ the one-band approximation is  justified.

\vspace{0.2 in}
 
\textbf{III. The single band model}
 
\vspace{0.2 in}
As a result of the discussion in the previous chapter we conclude that  for a finite chemical potential the  surface of the $TI$ can be described by  a single band with a correlated potential. 
\begin{equation}
H^{(c)}=\sum_{\vec{k}}(\hbar v|\vec{k}|-\mu)C^{\dagger}(\vec{k})C(\vec{k})+\sum_{\vec{k}}\sum_{\vec{p}} v U_{sc}(\vec{k}-\vec{p}) V^{(p,p)}(\vec{k}, \vec{p})C^{\dagger}(\vec{k})C(\vec{p})
\label{op}
\end{equation}
Where $V^{(p,p)}(\vec{k}, \vec{p})\equiv \langle u^{(+)}(\vec{k})|u^{(+)}(\vec{p})\rangle$ is evaluated  with  the help of  explicit form of eigenspinors   given in equation $(4)$. We introduce new operators $ \hat{C}^{\dagger}(\vec{k})$  and  $\hat{C}(\vec{k})$:
\begin{eqnarray}
&&\hat{C}^{\dagger}(\vec{k})= C^{\dagger}(\vec{k}) \Big[e^{\frac{i}{2}\chi(\vec{k})}\theta[M(\vec{k})]+e^{\frac{-i}{2}\chi(\vec{k})}\theta[-M(\vec{k})]\Big]\nonumber\\&&
\hat{C}(\vec{k})=\Big[e^{\frac{-i}{2}\chi(\vec{k})}\theta[M(\vec{k})]+e^{\frac{i}{2}\chi(\vec{k})}\theta[-M(\vec{k})]\Big]C(\vec{k})\nonumber\\&&
\end{eqnarray}
As a result we can express the Hamiltonian in eq. $(14)$ in the following way:
\begin{equation}
H^{(c)}=\sum_{\vec{k}}(\hbar v|\vec{k}|-\mu) \hat{C}^{\dagger}(\vec{k})\hat{C}(\vec{k})+ \sum_{\vec{k}}\sum_{\vec{p}}v \hat{C}^{\dagger}(\vec{k})\Gamma_{\|}(\vec{k},\vec{p})\hat{C}(\vec{p})
\label{ham} 
\end{equation}
In eq. ($16$)  $\Gamma_{\|}(\vec{k},\vec{p})$ represents the   scattering matrix.
\begin{equation}
\Gamma_{\|}(\vec{k},\vec{p})\equiv U_{sc}(\vec{k}-\vec{p})\hat{\Gamma}_{\|}(\vec{k},\vec{p});\hspace{0.1 in}
\hat{\Gamma}_{\|}(\vec{k},\vec{p})=\cos[\frac{1}{2}(\chi(\vec{k})-\chi(\vec{p}))]
\label{matrix}
\end{equation}
The angle $\chi(\vec{p})$  acts as a half  vortex, when the momentum is reversed the angle  $\chi(\vec{p})$ increases  by $\pi$, $\chi(-\vec{p})= \chi(\vec{p})+\pi$. As a result  we obtain:
 $\hat{\Gamma}_{\|}(\vec{k},-\vec{p})=\cos[\frac{1}{2}(\chi(\vec{k})-\chi(-\vec{p}))]=\cos[\frac{1}{2}(\chi(\vec{k})-\chi(\vec{p}))-\frac{\pi}{2}]= \sin[\frac{1}{2}(\chi(\vec{k})-\chi(\vec{p}))]$
 
\noindent For the \textbf{backscattering} direction  $\vec{p}=-\vec{k}$ we find that \textbf{the single particle scattering vanishes} $\hat{\Gamma}_{\|}(\vec{k},-\vec{k})=0$.

Next we consider  the  coupling of the external vector potential to the conduction band only:
\begin{eqnarray}
&&H^{ext}(t)=\int\ d^2 r[ J_{1}(\vec{r},t)A_{1}(\vec{r},t) + J_{2}(\vec{r},t)A_{2}(\vec{r},t)]=(-ev)\int\frac{d^2k}{(2\pi)^2}\int\frac{d^2p}{(2\pi)^2}\nonumber\\&&C^{\dagger}(\vec{k},t)C(\vec{p},t)[ \langle u^{(+)}(\vec{k})|-\sigma^2 |u^{(+)}(\vec{p})\rangle A_{1}(-(\vec{k}-\vec{p}),t)+\langle u^{(+)}(\vec{k})|\sigma^1 |u^{(+)}(\vec{p})\rangle A_{2}(-(\vec{k}-\vec{p}),t)]\nonumber\\&&=(-ev)\int\frac{d^2k}{(2\pi)^2}\int\frac{d^2p}{(2\pi)^2}\hat{C}^{\dagger}(\vec{k},t)\hat{C}(\vec{p},t)\Big[W_{1}(\vec{k},\vec{p})A_{1}(-(\vec{k}-\vec{p}),t)  +W_{2}(\vec{k},\vec{p})A_{2}(-(\vec{k}-\vec{p}),t)\Big] , \nonumber\\&&
\end{eqnarray}
where the vertex functions are given by:
\begin{equation}
W_{1}(\vec{k},\vec{p})=\cos[\frac{1}{2}(\chi(\vec{k})+\chi(\vec{p}))];\hspace{0.1 in} W_{2}(\vec{k},\vec{p})=-\sin[\frac{1}{2}(\chi(\vec{k})+\chi(\vec{p}))].
\label{eqphot}
\end{equation}
Using the definition of the current operator 
 $J_{1}(\vec{q},t)  =(-ev)\int\frac{d^{2}k}{(2\pi)^2}\int\frac{d^{2}p}{(2\pi)^2}W_{1}(\vec{k},\vec{p})\hat{C}^{\dagger}(\vec{k},t)\hat{C}(\vec{p},t)\delta^{2}[\vec{k}-\vec{p}-\vec{q}]$
 we obtain from the linear response theory the induced current
 $ \delta \langle J_{1}(\vec{q},t)\rangle=\int_{-\infty}^{\infty}dt_{1}R_{1,1}(\vec{q},t-t_{1})A_{1}(-\vec{q},t_{1})$
 with the  correlation function  $R_{1,1}(\vec{q},t-t_{1})$  given by \cite{Doniach}:
\begin{equation}
R_{1,1}(\vec{q},t-t_{1})=-\frac{i}{\hbar}\theta[t-t_{1}]\langle\Big[J_{1}(\vec{q},t),J_{1}(-\vec{q},t_{1})\Big]\rangle . 
\label{cor}
\end{equation}
We perform the Fourier transform of $R_{1,1}(\vec{q},t-t_{1})$  with respect the frequency $\Omega$ and find that the conductivity  is given by  $\sigma(\vec{q}=0,\Omega)\equiv \frac{R_{1,1}(\vec{q}=0,\Omega)}{i\Omega}= \sigma(\Omega)$.

\vspace{0.2 in}
 
\textbf{IV. Interference effects and the consequence for the conductivity}
 
\vspace{0.2 in}

The quantum conductivity is controlled by the  multiple scattering process of closed paths. The  interference  between a closed path and  the  time reversed paths  allows to define the quantum return probability  $P_{qm}$.
If we denote by $\mathscr{T}$  the amplitude for a particle to return to the starting point and  $T\mathscr{T}$   the time reversed amplitude, the  quantum return probability  $P_{qm}$ is given by   $P_{qm}= (\mathscr{T}+T\mathscr{T})(\mathscr{T}+T\mathscr{T})^{*}$  ($T$is the time reversal operator).
For   \textbf{time reversal invariant} systems  the quantum return probability  is $P_{qm}=4 (\mathscr{T})(\mathscr{T})^{*}$  and for spin half systems with spin-orbit interactions $P_{qm}= (\mathscr{T})(\mathscr{T})^{*}$ \cite{Bergmann}. Since the classical return probability  $P_{cl}= 2(\mathscr{T})(\mathscr{T})^{*}$ is smaller  for time reversal invariant and larger for spin half systems with spin-orbit interactions one conclude that quantum interference gives rise to weak localization    for the first case and anti-localization for the second   case.

The  scattering  amplitude  to order $n+1$ for a particle with momentum $\vec{k}$ to be  backscattering to $-\vec{k}$  is  is given by $\mathscr{T}(\vec{k},-\vec{k})$:
\begin{eqnarray}
&&\mathscr{T}(\vec{k},-\vec{k})=  \Gamma_{\|}(\vec{k},-\vec{k})+\sum_{\vec{p}}\frac{\Gamma_{\|}(\vec{k},\vec{p})\Gamma_{\|}(\vec{p},-\vec{k})}{E(\vec{k})-E(\vec{p})+i\epsilon}+\sum_{\vec{p}}\sum_{\vec{l}}\frac{\Gamma_{\|}(\vec{k},\vec{p})\Gamma_{\|}(\vec{p},\vec{l})\Gamma_{\|}(\vec{l},-\vec{k})}{(E(\vec{k})-E(\vec{p})+i\epsilon)(E(\vec{k})-E(\vec{l})+i\epsilon)}+\nonumber\\&&
..\sum_{\vec{p}_{1}}\frac{\Gamma_{\|}(\vec{k},\vec{p}_{1})}{E(\vec{k})-E(\vec{p}_{1})+i\epsilon}\Big[\sum_{\vec{p}_{2}}  ..\sum_{\vec{p}_{n}}\frac{\Gamma_{\|}(\vec{p}_{1},\vec{p}_{2})...\Gamma_{\|}(\vec{p}_{n-1},\vec{p}_{n})}{(E(\vec{k})-E(\vec{p}_{2})+i\epsilon)..(E(\vec{k})-E(\vec{p}_{n-1})+i\epsilon)}\Big]\frac{\Gamma_{\|}(\vec{p}_{n},-\vec{k})}{E(\vec{k})-E(\vec{p}_{n})+i\epsilon}\nonumber\\&&
\end{eqnarray}
From equation $(21)$ we observe  that the first order  backscattering term  is zero $\Gamma_{\|}(\vec{k},-\vec{k})=0$.
The time reversal amplitude   $T \mathscr{T}(\vec{k},-\vec{k})$ is obtained by reversing all the momenta
\begin{eqnarray}
&&T \mathscr{T}(\vec{k},-\vec{k})\equiv  \mathscr{T}(-\vec{k},\vec{k})= \Gamma_{\|}(-\vec{k},\vec{k}) +\sum_{\vec{p}}\frac{\Gamma_{\|}(-\vec{k},-\vec{p})\Gamma_{\|}(-\vec{p},\vec{k})}{E(-\vec{k}-E(-\vec{p})+i\epsilon}+\nonumber\\&&
\sum_{\vec{p}}\sum_{\vec{l}}\frac{\Gamma_{\|}(-\vec{k},-\vec{p})\Gamma_{\|}(-\vec{p},-\vec{l})\Gamma_{\|}(-\vec{l},\vec{k})}{(E(-\vec{k})-E(-\vec{p})+i\epsilon)(E(-\vec{k})-E(-\vec{l})+i\epsilon)}+..\sum_{\vec{p}_{1}}\frac{\Gamma_{\|}(-\vec{k},-\vec{p}_{1})}{E(\vec{k})-E(\vec{p}_{1})+i\epsilon}\cdot\nonumber\\&&
\Big[\sum_{\vec{p}_{2}}  \sum_{\vec{p}_{n}}\frac{\Gamma_{\|}(-\vec{p}_{1},-\vec{p}_{2})...\Gamma_{\|}(-\vec{p}_{n-1},-\vec{p}_{n})}{(E(-\vec{k})-E(-\vec{p}_{2})+i\epsilon)..(E(-\vec{k})-E(-\vec{p}_{n-1})+i\epsilon)}\Big]\frac{\Gamma_{\|}(-\vec{p}_{n},\vec{k})}{E(-\vec{k})-E(-\vec{p}_{n})+i\epsilon}\nonumber\\&&
\end{eqnarray}
Using the fact that the eigenvalues obey $E(-\vec{p})=E(\vec{p})$ and the scattering vertex satisfies the relation $\Gamma_{\|}(-\vec{p}_{1},-\vec{p}_{2})=\Gamma_{\|}(\vec{p}_{1},\vec{p}_{2})$ we find that difference between the two amplitudes comes from the terms outside the bracket.
Using the relation $\chi(-\vec{k})= \chi(\vec{k})+\pi$ we obtain  :
\begin{equation}
\Gamma_{\|}(\vec{k},\vec{p})\Gamma_{\|}(\vec{p}_{n},-\vec{k})=-\Gamma_{\|}(-\vec{k},-\vec{p})\Gamma_{\|}(-\vec{p}_{n},\vec{k})
\label{identity}
\end{equation}
As a result  $|(\mathscr{T}(\vec{k},-\vec{k})+ T\mathscr{T}(\vec{k},-\vec{k}))|=0$ and therefore $P_{qm}=0$.
Next we compute compute the amplitude  for the scattering of a particle with incoming  momentum $\vec{k}$ to  outgoing momentum  $-\vec{k}=\vec{Q}$:
$P_{qm}(\vec{Q})=(\mathscr{T}(\vec{k},-\vec{k}+\vec{Q})+T\mathscr{T}(\vec{k},-\vec{k}+\vec{Q}))(\mathscr{T}(\vec{k},-\vec{k}+\vec{Q})+T\mathscr{T}(\vec{k},-\vec{k}+\vec{Q}))^*$
Using the properties of the   vertex functions:
\begin{eqnarray}
&&\hat{\Gamma}_{\|}(\vec{k},\vec{p})\hat{\Gamma}_{\|}(\vec{p},-\vec{k}+\vec{Q}) \hat{\Gamma}_{\|}(-\vec{k},-\vec{p})\hat{\Gamma}_{\|}(-\vec{p},\vec{k}-\vec{Q})\approx-\nonumber\\&& \frac{1}{4}\sin^2[\chi(\vec{k})-\chi(\vec{p})]\cos[\frac{1}{2}(\chi(\vec{k})-\chi(\vec{k}-\vec{Q}))]\cos[\frac{1}{2}(\chi(\vec{p})-\chi(\vec{p}-\vec{Q}))]\cos[\frac{1}{2}(\chi(\vec{k}-\vec{Q})-\chi(\vec{p}-\vec{Q}))]\nonumber\\&& \approx  -\frac{1}{4}\sin^2[(\chi(\vec{k})-\chi(\vec{p}))][1-\frac{ \vec{Q}^2}{2k^2_{F}}]\nonumber\\&&
\end{eqnarray}
we compute the amplitude for small angle backscattering: 

$P_{qm}(\vec{Q})=(\mathscr{T}(\vec{k},-\vec{k}+\vec{Q})+T\mathscr{T}(\vec{k},-\vec{k}+\vec{Q}))(\mathscr{T}(\vec{k},-\vec{k}+\vec{Q})+T\mathscr{T}(\vec{k},-\vec{k}+\vec{Q}))^*$
Following the analysis performed  for  the backscattering result obtained for  scalar waves in    time reversed  systems  \cite{kaveh} we compute 
 the backscattering amplitude at small angle (small $\vec{Q}$). For incoming electrons with momentum $\vec{k}$  $|\vec{k}|\approx k_{F}$ which are backscattered at an angle $\theta$ we introduce $k_{F}\sin[\theta]=|\vec{Q}|$ and  find in terms of the    the elastic mean free path $l$ \cite{kaveh} :
\begin{equation}
P_{qm}(\vec{Q})=(\mathscr{T}(\vec{k},-\vec{k}+\vec{Q})+T\mathscr{T}(\vec{k},-\vec{k}+\vec{Q}))(\mathscr{T}(\vec{k},-\vec{k}+\vec{Q})+T\mathscr{T}(\vec{k},-\vec{k}+\vec{Q}))^*\approx 1-\frac{1}{(1+k_{F}l \sin[\theta])^2}
\label{sol}
\end{equation}
It is important to stress that this result has been obtained for a time reversal invariant system in the presence of a half  vortex  at $\vec{k}=0$.

\vspace{0.2 in}
 
\textbf{V. The Green's function  for spinless electrons in the presence of disorder}
 
\vspace{0.2 in}

In order to compute the conductivity we need to include    the interference results established in section $IV$.  This  will  be done using  the method of Green's functions for a random potential described in the literature  \cite{Doniach}.
We will use the Green's function  $G(\vec{k},s;\vec{p},0)$ for the conduction  band in the presence of the random potential   and  $G^{(0)}(\vec{k},s)$,  the Green's function in the absence of disorder:
\begin{eqnarray}
&&G(\vec{k},s;\vec{p},0)=-i\langle T(\hat{C}(\vec{k},s)\hat{C}^{\dagger}(\vec{p},0)\rangle , \nonumber\\&& G^{(0)}(\vec{k},s)=-i\langle 0|T(\hat{C}(\vec{k},s)\hat{C}^{\dagger}(\vec{k},0)|0\rangle , \nonumber\\&&
G^{(0)}(\vec{k},\omega)=\int_{-\infty}^{\infty}\,ds e^{i\omega s}G^{(0)}(\vec{k},s)=[\hat{\omega} -E(\vec{k}) +i\epsilon sgn (\hat{\omega})]^{-1}=[\omega -\epsilon(\vec{k}) +i\epsilon sgn (\omega)]^{-1} \nonumber\\&&
\end{eqnarray}
where $\hat{\omega}=\omega-\mu$ and $\epsilon(\vec{k})\equiv E(\vec{k})-\mu$ is the energy measured with respect the chemical potential $\mu>0$.

\textbf{a-The Averaged Single Particle Green's Function}

Using the  free Green's functions $G^{(0)}(\vec{k},\omega) $ we compute the self energy as a function of the statistical average of the  disorder  potential. 
\begin{eqnarray}
&&\Sigma(\vec{k},\omega)=v^{2}\int\,\frac{d^2 p }{(2\pi)^2}\langle\langle U_{sc}(\vec{k}-\vec{p})  U_{sc}(\vec{p}-\vec{k}) \rangle\rangle \hat{\Gamma}_{\|}(\vec{k},\vec{p})\hat{\Gamma}_{\|}(\vec{p},\vec{k})G^{(0)}(\vec{p},\omega) \nonumber\\&&
= D_{sc}v^{2}\int\,\frac{d^2 p }{(2\pi)^2}\cos^2[\frac{1}{2}(\chi(\vec{k})-\chi(\vec{p})][\omega-\epsilon(\vec{p})+i\epsilon sgn(\omega)]^{-1} , \nonumber\\&&
\end{eqnarray}
We replace the integration with respect $\chi$ by a polar angle integration on the Fermi surface $E(\vec{k})=\mu$:

$\frac{d^2 p }{(2\pi)^2}=\frac{pdp}{2\pi}\frac{d\theta}{2\pi}=\frac{(\mu+\epsilon)d\epsilon}{2\pi v^2}(\frac{d\chi(\vec{p}}{d\theta})^{-1}|_{\vec{p}=\vec{k_{F}}} \frac{d\chi(\vec{p})}{2\pi}\approx \frac{(\mu+\epsilon)d\epsilon}{2\pi v^2}\frac{d\chi(\vec{p})}{2\pi}$

\noindent  We obtain for the self energy:
\begin{equation}
\Sigma(\vec{k},\omega)=D_{sc}\int_{-\mu}^{v \hbar \Lambda-\mu}\frac{(\mu+\epsilon)d \epsilon}{2\pi v^2}\int \frac{d\chi(\vec{p})}{2\pi}\cos[\frac{1}{2}(\chi(\vec{k})-\chi(\vec{k}))][\omega-\epsilon(\vec{p})+i\epsilon sgn(\omega)]^{-1}
\label{self}
\end{equation}
\begin{eqnarray}
&&\Sigma(\vec{k},\omega)\equiv \Sigma_{R}(\vec{k},\omega)+i \Sigma_{I}(\vec{k},\omega) , \nonumber\\&&
\Sigma_{I}(\vec{k},\omega)=\pi D_{sc}\frac{1}{2}\int_{-\mu}^{v \hbar \Lambda-\mu}\frac{(\mu+\epsilon)d\epsilon}{2\pi}\delta[\omega-\epsilon]=\frac{D_{sc}\mu}{4}(1+\frac{\hbar\omega}{\mu})\equiv \frac{1}{2\tau}\nonumber\\&&
\Sigma_{R}(\vec{k},\omega)=-\frac{D_{sc}}{4 \pi} (\mu+\hbar\omega)\log(\frac{v\hbar\Lambda-\mu-\hbar \omega}{\hbar \omega+\mu})-v\hbar\Lambda\nonumber\\&&
E(\vec{p})-\mu=v\hbar|\vec{p}|=\epsilon(\vec{k}); \mu=v\hbar k_{F}
\nonumber\\&&
\end{eqnarray}
The  real part of the self energy  allows to compute the wave function renormalization  $Z$. The value $Z$  is determined by the band width cut-off $\Lambda $, the chemical potential $\mu$ and the elastic mean free path $l$:
\begin{eqnarray}
&&Z =[1-\partial_{\hat{\omega}}\Sigma_{R}(\hat{\omega},\mu)|_{\hat{\omega}=0}]=\Big[1-\frac{1}{2\pi k_{F}l}[1+\frac{1}{\frac{v\hbar\Lambda}{\mu}-1}-Log(\frac{1}{\frac{v\hbar\Lambda}{\mu}-1})]\Big]^{-1}\nonumber\\&&
\end{eqnarray}
Considering the experimental situation we have $\frac{\hbar v \Lambda}{\mu}\approx 3$  and $k_{F}l>1 $ the wave function renormalization  $Z\approx 1$.

The averaged Green's function  $\mathbb{G}(\vec{p},\omega)$  is  expressed in terms of the renormalized wave function $Z$,   renormalized velocity $v_{R}$ and renormalized life time $\tau_{R}$. We find:
\begin{equation}
\mathbb{G}(\vec{p},\omega)=\frac{Z}{\omega-\epsilon_{R}(\vec{p})+i(\frac{1}{2\tau _{R}})sgn(\omega)},\hspace{0.1 in} \epsilon_{R}(\vec{p})=\hbar \vec{v}_{R}\cdot \vec{p},\hspace{0.1 in} v_{R}=vZ,\hspace{0.1 in} \tau_{R}=\tau Z^{-1} . 
\label{ez}
\end{equation}
Here $\mathbb{G}_{+,R}(\epsilon_{R}(\vec{p}),\omega)=\frac{Z}{\omega -\epsilon_{R}(\vec{p})+i\frac{Z}{2\tau}}$ is the retarted  Green's function and   $\mathbb{G}_{-,R}(\epsilon_{R}(\vec{p}),\omega)=\frac{Z}{\omega -\epsilon_{R}(\vec{p})-i\frac{Z}{2\tau}}$ is the advanced Green's function.

\textbf{b-The ladder diagram-renormalized charge and transport scattering time}

Next we compute the effect of the ladder diagrams.
The ladder diagrams   renormalized  the charge $e$ and causes the replacement of the elastic scattering time $\tau$ with the transport time $\tau^{tr.}$.
The calculation will be done at zero momentum. We introduce the renormalized charge  $e_{R}(\Omega)$ at a finite frequency  $\Omega$. The current is determined by the product $ev$ and the photon  vertex $W_{1}(\vec{k},\vec{k})$.
In our case the velocity normalizes separately such that the product $vC^{\dagger}(\vec{k})C(\vec{k})=v_{R}C_{R}^{\dagger}(\vec{k})C_{R}(\vec{k})$ is invariant.  The renormalized charge satisfies  the following integral equation.
\begin{eqnarray}
&&e_{R}(\Omega)v W_{1}(\vec{k},\vec{k})=e v W_{1}(\vec{k},\vec{k})+e_{R}(\Omega)v \int\,\frac{d^2p}{(2\pi)^2}  W_{1}(\vec{p},\vec{p}) [\langle\langle U_{sc}(\vec{k}-\vec{p})U_{sc}(\vec{}-\vec{k})\rangle\rangle] v^2\hat{\Gamma}_{\|}(\vec{k},\vec{p})\hat{\Gamma}_{\|}(\vec{p},\vec{k})\nonumber\\&&
\mathbb{G}_{+,R}(\epsilon_{R}(\vec{p}),\omega)\mathbb{G}_{-,R}(\epsilon_{R}(\vec{p}),\omega+\Omega) 
\nonumber\\&&
\end{eqnarray}
In order to solve the integral equation we use the explicit form of the scattering vertex product, $\hat{\Gamma}_{\|}(\vec{k},\vec{p})\hat{\Gamma}_{\|}(\vec{p},\vec{k})=\cos^2[\frac{1}{2}(\chi(\vec{k})-\chi(\vec{p}))]$. In addition we replace the photon vertex $W_{1}(\vec{p},\vec{p})=\cos[\chi(\vec{p})]$ by 
$\cos[\chi(\vec{p})]=\cos[\chi(\vec{k})+(\chi(\vec{p})-\chi(\vec{k}))]$, using trigonometric relations  we  find: 
\begin{equation}
\int\frac{d\chi(\vec{p})}{2\pi}  W_{1}(\vec{p},\vec{p})\hat{\Gamma}_{\|}(\vec{k},\vec{p})\hat{\Gamma}_{\|}(\vec{p},\vec{k})=W_{1}(\vec{k},\vec{k}) \int\frac{d\chi(\vec{p})}{2\pi} \cos[\chi(\vec{p})-\vec{p})] \cos^2[\frac{1}{2}(\chi(\vec{k})-\chi(\vec{p}))]=\frac{1}{4} W_{1}(\vec{k},\vec{k})
\label{integral}
\end{equation}
As a result  we factor out  $W_{1}(\vec{k},\vec{k})$   on both side of the integral equation.
We perform the disorder average $[\langle\langle U_{sc}(\vec{k}-\vec{p})U_{sc}(\vec{}-\vec{k})\rangle\rangle]=D_{sc}$,  introduce the notation for $\frac{I(\Omega)D_{sc}}{2}$  and   and perform the  energy $\epsilon(\vec{p})$ integration:
\begin{equation}
\frac{I(\Omega)D_{sc}}{2}= \frac{D_{sc}\mu}{2}\int \frac{d\epsilon(\vec{p})}{2\pi}\mathbb{G}_{+,R}(\epsilon_{R}(\vec{p}),\omega)\mathbb{G}_{-,R}(\epsilon_{R}(\vec{p}),\omega+\Omega)\approx \frac{0.5 D_{sc}\mu \tau}{1+i \Omega \tau Z^{-1}}
\label{eqv}
\end{equation}
Using this result we obtain the  solution for  the integral equation:
\begin{eqnarray}
&&e_{R}(\Omega)=\frac{e}{1-\frac{I(\Omega)D_{sc}}{4}}=\frac{e}{1-\frac{1}{2}\frac{1}{1+i \Omega \tau Z^{-1}}}\nonumber\\&&
e_{R}(\Omega\rightarrow 0) \rightarrow  2 e  \nonumber\\&&
\end{eqnarray} 
The enhancement of the charge  by a factor of $2$ can be interpreted as increase of the  transport time  with respect the life time $\tau^{tr.}=4\tau$.

\textbf{c-The contribution of the maximal crossed diagrams}

Following \cite{Rammer} we compute the maximal crossed diagram   for the  conduction  electrons with a fixed chirality   using the scattering vertex  $\Gamma_{\|}(\vec{k},\vec{p})$.  Following the discussion given in section $III$ we obtain that the maximal crossed will become negative due to the identity $\chi(-\vec{p})= \chi(\vec{p})+\pi$.
We introduce the notation for  two particles  $Cooperon$ scattering: 
\begin{equation}
\mathscr{C}[ -\vec{p}+\vec{Q},\vec{p}; -\vec{k}+\vec{Q},\vec{k};\Omega]\equiv \mathbb{C}[\vec{p},\vec{k};\vec{Q},\Omega]
\label{coop}
\end{equation}
where $-\vec{k}+\vec{Q},\vec{k}$ are the momenta of the incoming  particles and $ -\vec{p}+\vec{Q},\vec{p}$ are the momenta of the  outgoing particles. In the second  notation $\vec{Q}$  represent the total momentum conserved in the process. We will work with the short notation $\mathbb{C}[\vec{p},\vec{k};\vec{Q},\Omega]$. We replace  $[\langle\langle U_{sc}(\vec{k}-\vec{p})U_{sc}(\vec{p}-\vec{k})\rangle\rangle]$ by the statistical average $D_{sc}$ and obtain the integral equation:
\begin{eqnarray}
&&\mathbb{C}[\vec{k},\vec{k};\vec{Q},\Omega]=D^2_{sc}\int\,\frac{d^2 p}{(2\pi)^2}\nonumber\\&&\hat{\Gamma}_{\|}(\vec{k},-\vec{p})\hat{\Gamma}_{\|}(-\vec{p},-\vec{k}+\vec{Q})
\hat{\Gamma}_{\|}(-\vec{k}+\vec{Q},\vec{p}+\vec{Q})\hat{\Gamma}_{\|}(\vec{p}+\vec{Q},\vec{k})\mathbb{G}_{+,R}(-\vec{p},\omega)\mathbb{G}_{-,R}(\vec{p}+\vec{Q},\omega+\Omega)\nonumber\\&&
+D_{sc}\int\,\frac{d^2 p}{(2\pi)^2}\mathbb{C}[\vec{k},\vec{p};\vec{Q},\Omega]       \mathbb{G}_{+,R}(-\vec{p},\omega)\mathbb{G}_{-,R}(\vec{p}+\vec{Q},\omega+\Omega)\hat{\Gamma}_{\|}(\vec{k},\vec{p})\hat{\Gamma}_{\|}(-\vec{k}+\vec{Q},-\vec{p})\nonumber\\&&
\end{eqnarray}
$\mathbb{C}[\vec{k},\vec{k};\vec{Q},\Omega]$ is negative!
In particular we observe that the $n$ order term is proportional to:
\begin{equation}
-\frac{1}{4}\sin^2[\chi(\vec{k})-\chi(\vec{p}_{1})]\prod_{i=1}^{n-2}\cos^2[\frac{1}{2}(\chi(\vec{p}_{i})-\chi(\vec{p}_{i+1})]
\label{angular}
\end{equation}
We solve iteratively the integral equation  and perform the  integration with respect the  variables  $\chi(\vec{p}_{1})$,$\chi(\vec{p}_{2})$,..$\chi(\vec{p}_{n-1})$.
We introduce the notation:
\begin{equation}
\frac{I(\Omega,\vec{Q})D_{sc}}{2}= \frac{D_{sc}\mu}{2}\int \frac{d\epsilon(\vec{p})}{2\pi}\mathbb{G}_{+,R}(\epsilon_{R}(\vec{p}),\omega)\mathbb{G}_{-,R}(\epsilon_{R}(\vec{p}+\vec{Q}),\omega+\Omega)\approx \frac{\frac{1}{2} D_{sc}\mu \tau}{1+i \Omega \tau Z^{-1}- \frac{1}{2}l^2Q^2}
\label{eqv}
\end{equation}
where $l\equiv v\tau= v_{R}\tau_{R}$. We find that $\mathbb{C}[\vec{k},\vec{k};\vec{Q},\Omega]$  is given by:
\begin{eqnarray}
&&\mathbb{C}[\vec{k},\vec{k};\vec{Q},\Omega]=-\frac{1}{4}D_{sc}\Big[\frac{I(\Omega,\vec{Q})D_{sc}}{2}+...\Big[\frac{I(\Omega,\vec{Q})D_{sc}}{2}\Big]^{n}...\Big]\nonumber\\&&=\frac{1}{4}D_{sc}\Big[1-\frac{1}{1-\frac{I(\Omega,\vec{Q})D_{sc}}{2}}\Big]\nonumber\\&&
\end{eqnarray}
This equation has the same form as we have shown  in equation $(24)$. Next we substitute  the explicit result for  
$\frac{I(\Omega,\vec{Q})D_{sc}}{2}$ given in eq.$(39)$ and find:
\begin{equation}
\mathbb{C}[\vec{k},\vec{k};\vec{Q},\Omega]=\frac{1}{4}D_{sc}\Big[1-\frac{1}{-i\Omega\tau Z^{-1}+\frac{1}{2}l^2Q^2}\Big]
\label{intecooperon}
\end{equation}
This result shows that the Cooperon has a \textbf{negative} sign therefore we will have  anti-localization.

\vspace{0.2 in}
 
\textbf{VI  Computation of the conductivity}
 
\vspace{0.2 in}

In this section we will compute the conductivity $\sigma(\vec{q}=0,\Omega \rightarrow 0) $ using the maximal crossed diagrams.  We wind that the conductivity is determined by the   longitudinal  correlation function  $R_{1,1}(\vec{q},t-t_{1})$: 
\begin{eqnarray}
&&R_{1,1}(\vec{q},t-t_{1})=-\frac{i}{\hbar}\theta[t-t_{1}]\langle\Big[J_{1}(\vec{q},t),J_{1}(-\vec{q},t_{1})\Big]\rangle\nonumber\\&&=-\frac{i}{\hbar}\theta[t-t_{1}](-e v)^{2}\int\,\frac{d^2 k }{(2\pi)^2}\int\,\frac{d^2 p }{(2\pi)^2}\int\,\frac{d^2 k_{1} }{(2\pi)^2}\int\,\frac{d^2 p_{1} }{(2\pi)^2}\nonumber\\&&W_{1}(\vec{k},\vec{p})W_{1}(\vec{k}_{1},\vec{p}_{1})\delta ^{(2)}[\vec{k}-\vec{p}-\vec{q}]\delta ^{(2)}[\vec{k}_{1}-\vec{p}_{1}+\vec{q}]\langle\Big[\hat{C}^{\dagger}(\vec{k},t))\hat{C}(\vec{p},t)),\hat{C}^{\dagger}(\vec{k}_{1},t_{1}))\hat{C}(\vec{p}_{1},t_{1})\Big]\rangle . \nonumber\\&&
\end{eqnarray}
We perform the Fourier transform   with respect to frequency $\Omega$ and find that the conductivity  is given by  $\sigma(\vec{q}=0,\Omega)\equiv \frac{R_{1,1}(\vec{q}=0,\Omega)}{i\Omega}= \sigma(\Omega)$.
Using the  averaged Green's function  over the random potential we obtain:
\begin{eqnarray}
&&\sigma(\Omega)=\frac{-i(ev)^2}{i\Omega \hbar}\int\,\frac{d^2 p}{(2\pi)^2}\int\,\frac{d^2 p_{1}}{(2\pi)^2}W_{1}(\vec{p},\vec{p})W_{1}(\vec{p}_{1},\vec{p}_{1})\int_{-\infty}^{\infty}\,ds e^{i\Omega s}\nonumber\\&&\Big[\theta[s]\overline{G(\vec{p},\vec{p}_{1};s)G(\vec{p}_{1},\vec{p};-s)}-\theta[s](\overline{G(\vec{p},\vec{p}_{1};-s)G(\vec{p}_{1},\vec{p};s)})^{*}\Big] , \nonumber\\&&
\sigma(\Omega)=\sigma^{(0)}(\Omega)+\sigma^{(1)}(\Omega) .\nonumber\\  
\end{eqnarray}
Next we introduce  $\mathbb{G}_{+}$ and $\mathbb{G}_{-}$, the retarded and advanced  averaged Green's functions \cite{Abrikosov}. $\sigma^{(0)}(\Omega)$  represents the real part of the  conductivity, computed with the averaged single particle Green's function $\overline{G}(\vec{p},\omega)$ and the ladder effect will be included  with the help of the  renormalized charge $e_{R}$ given in eq. $(34)$. $\sigma^{(1)}(\Omega)$ represents the correction to the conductivity obtained from  the maximal crossed diagrams \cite{Rammer}.  $\sigma^{(0)}(\Omega)$ is given by:
\begin{eqnarray}
&&\sigma^{(0)}(\Omega)=-\frac{e_{R}^2 \mu_{R}}{\hbar \Omega}\int\frac{d \chi}{2\pi}(\cos^{2}[\chi])\int_{-\epsilon_{min}}^{\epsilon_{max}} \frac{d\epsilon_{R}}{2\pi}\int_{-\Omega}^{0}\,\frac{d\omega}{2\pi}2\mathbb{G}_{+,R}(\epsilon_{R},\omega)\mathbb{G}_{-,R}(\epsilon_{R},\omega+\Omega)\nonumber\\&&
=-\frac{e_{R}^{2} \mu_{R}}{2h}\left(\frac{-i}{-\Omega - \frac{i}{\tau_{R}}}\right)
=\frac{e^{2}_{R}(\Omega)}{2h}k_{F}l\left(\frac{1-i(\Omega \tau)Z^{-1}}{1+(\Omega \tau)^2 Z^{-2}}\right) . \nonumber\\&&
\end{eqnarray} 
In order to compute the quantum corrections we need to compute the maximal crossed  diagram.
Using the maximal crossed diagram we consider the quantum correction to the conductivity following the so called Hikami boxes  \cite{Hikami}. There are three such diagrams (see figure 4 in ref. \cite{Hikami},
The leading contribution to the conductivity is given by:
\begin{eqnarray}
&&\sigma^{(1)}(\Omega\rightarrow 0)=
-\frac{i (e_{R} v)^2}{\hbar i\Omega}2 \int\,\frac{d^2 p}{(2\pi)^2}\int\,\frac{d^2 Q}{(2\pi)^2}\int_{-\Omega}^{0}\,\frac{d\omega}{2\pi}W_{1}(\vec{p},\vec{p})W_{1}(-\vec{p}+\vec{Q},-\vec{p}+\vec{Q})\nonumber\\&&\mathbb{G}_{+,R}(\vec{p},\omega)\mathbb{G}_{-,R}(\vec{p},\omega)\mathbb{G}_{+,R}(-\vec{p}+\vec{Q},\omega)\mathbb{G}_{-,R}(-\vec{p}+\vec{Q},\omega)\Big[v\mathbb{C}[\vec{k},\vec{k};\vec{Q},\Omega=0]v\Big]\nonumber\\&&
\end{eqnarray}
Next we use the fact that the angular average in the limit $\vec{Q}=0$ is given by  $\langle W_{1}(\vec{p},\vec{p})W_{1}(-\vec{p}+\vec{Q},-\vec{p}+\vec{Q})\rangle=-\frac{1}{2}$.
Performing the momentum integration with the help of the residue theorem and neglecting the additive factor  $\frac{D_{sc}}{4}$ (see eq. $(41)$ ) gives:
\begin{equation}
\sigma^{(1)}(\Omega\rightarrow 0)=\frac{e_{R}^2}{h \pi}\int_{\frac{1}{L_{\phi}}}^{\frac{1}{l}}\frac{d Q}{Q} =
\frac{e_{R}^2}{h \pi} Ln \Big[\frac{L_{\phi}}{l}\Big]  
\label{corrections}
\end{equation}
This result shows  tat \textbf{anti-localization} as obtained. The corrections depend on the \textbf{phase-coherence length $L_{\phi}$ and the elastic scattering  length}. The phase-coherence length $L_{\phi}$ is limited by the length of the sample $L$, $L_{\phi}<L$.
The conductivity  is given as a sum of the two parts,$\sigma(\Omega=0)=\sigma^{(0)}(\Omega=0)+ \sigma^{(1)}(\Omega=0)$.  
\begin{equation}
\sigma(\Omega \rightarrow 0) \approx  \frac{4e^{2}}{2 h}k_{F}l\Big[1+ \frac{1}{\pi k_{F}l}Ln \Big[\frac{L_{\phi}}{l}\Big]\Big]
\label{log}  
\end{equation}
The real part of  the  conductivity as a function of the   frequency  $\frac{\Omega}{\Omega_{F}}$ , ( $\Omega_{F}=\frac{\mu}{\hbar}$), phase-coherence length   $\frac{L_{\phi}}{l}$  and renormalized charge  $e_{R}(\frac{\Omega}{\Omega_{F}})$ ) is  given by $(34)$.
\begin{equation}
\sigma(\Omega)=\frac{e^{2}_{R}(\frac{\Omega}{\Omega_{F}})}{2 h}k_{F}l\Big[\frac{1}{1+(\frac{\Omega}{\Omega_{F}}k_{F}l Z^{-1})^2}+\frac{1}{4\pi k_{F}l}Ln \Big[\frac{ (\frac{\Omega}{\Omega_{F}}k_{F}l Z^{-1})^2+\frac{1}{2}}{(\frac{\Omega}{\Omega_{F}}k_{F}l Z^{-1})^2 +\frac{1}{2}(\frac{l}{L_{\phi}})^2}\Big]\Big]
\label{equaf}
\end{equation}
In figure $(1)$ we show the conductivity  as a function of frequency  for $\frac{l}{L_{\phi}}=10^{-4}$ and $Z\approx 1$ for different values of $k_{F}l=1,2.5,5,10,20$. Figure $(1)$  shows that the conductivity increases with the decreases of $k_{F}l$   The figures also indicates that when  $\frac{l}{L_{\phi}}\rightarrow 0 $ and $\Omega=0$ the conductivity diverges. This results are in agreement with  with the anti-localization theory predicted by the symplectic ensemble \cite{Hikami}.

\begin{figure}
\begin{center}
\includegraphics[width=7.25 in ]{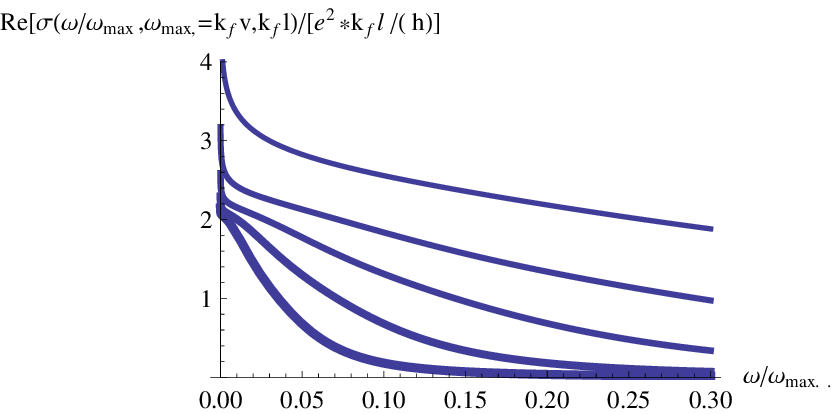}
\end{center}
\caption{The conductivity as a function of the frequency for three cases:  $k_{F}l=20$  (the lower graph), $k_{F}l=10$  $k_{F}l=5$ ,$k_{F}l=2.5$,  $k_{F}l=1$ (the upper graph). } 
\end{figure} 
 
\vspace{0.2 in}

\textbf{IV. Possible Experimental verification of the theory}

\vspace{0.2 in}

One of the difficulties in measuring the surface conductivity is the fact that the bulk electrons contribute to the conductivity. The bulk electrons form a three dimensional $TI$ characterized by the inverted gap $\Delta$. For the case that the chemical potential $\mu>\Delta$ the "` bulk"' $TI$ will form a conduction band. For this case the spinor will be given by $\Psi(\vec{r})=\sqrt{v}\sum_{\sigma=\uparrow,\downarrow}\sum_{\vec{k}}e^{i\vec{k}\cdot \vec{r}}C_{\sigma}(\vec{k})U_{\sigma}^{(+)}(\vec{k})$. In contrast to the surface case the particles carry two spin polarizations. As a result the Hamiltonian for this case will be replaced by: 
\begin{equation}
H^{(3D)}=\sum_{\sigma=\uparrow,\downarrow}\sum_{\vec{k}}(\hbar v(\sqrt{k^2 +\Delta^2})-\mu) \hat{C}_{\sigma}^{\dagger}(\vec{k})\hat{C}_{\sigma}(\vec{k})+ \sum_{\vec{k}}\sum_{\vec{p}}\sum_{\sigma=\uparrow,\downarrow} \sum_{\sigma'=\uparrow,\downarrow}v \hat{C}_{\sigma}^{\dagger}(\vec{k})\Gamma_{\sigma,\sigma'}(\vec{k},\vec{p})\hat{C}_{\sigma'}(\vec{p}) , 
\label{3D}
\end{equation}
where $\Gamma_{\sigma,\sigma'}(\vec{k},\vec{p})$ is the scattering matrix.
Using  a model of parallel resistors we find  that the conductivity will be dominated by the bulk contributions.
In order to observe the surface conductivity we need to be in a situation where $\mu<\Delta$. For this case the bulk will be insulating and the conductivity will be dominated by the surface part.
For this case electronic Raman scattering might be a good tool to observe the surface conductivity. This might work for  laser frequencies  $\omega_{L}$ which excite the surface but not the bulk $\Delta >\omega_{L}>\mu$.
In order to evaluate the Raman intensity  we integrate out the anti-particles and obtain the effective \textbf{two photon} Hamiltonian:
\begin{eqnarray}
&&H^{(ext.2)}=\int\frac{d^2 k}{(2\pi)^2}\int\frac{d^2 q}{(2\pi)^2}\int \frac{d^2 q'}{(2\pi)^2}\int\frac{d\omega}{2\pi}   
\Big[\frac{(ev)^{2} W^{(p,a)}_{i}(\vec{k}+\vec{q},-\vec{k})W^{(a,p)}_{j}(-\vec{k},\vec{k}-\vec{q}')}{(\omega- \mu-(E(\vec{k} +\mu))}\Big]\nonumber\\&& C^{\dagger}(\vec{k}+\vec{q},\omega+\omega_{L}+ \omega_{S})C(\vec{k}-\vec{q}',\omega)A_{i}(-\vec{q},-\omega_{L})A_{j}(-\vec{q}',-\omega_{s}) , \nonumber \\&&
\end{eqnarray}
where $A_{i}(-\vec{q},-\omega_{L})$ is the laser vector potential in the $i$ direction, $A_{j}(-\vec{q}',-\omega_{s})$
is the scattering laser frequency and $W^{(p,a)}_{i}(\vec{k}+\vec{q},-\vec{k})$ represent the photon  vertex  which couples to electrons and holes.
Taking the limit $\vec{q}\rightarrow 0$  we observe that the Raman frequency $\Omega=\omega_{L}-\omega_{S}$ can provide  information about the electronic conductivity.
The Raman spectrum  measured in \cite{Conder} is given by: 
$S(\Omega)=\frac{1}{\pi}[1+(e^{\beta\Omega}-1)^{-1}]Im.D[\Omega]$ where $D[\Omega]=\frac{-i}{L^{2}} \int_{0}^{\infty} \,dt e^{i(\Omega+i\epsilon)t}\langle[\rho(t),\rho(0)]\rangle$ represents the chiral  surface electronic  density-density correlation. Since $Im.D[\Omega]$ is proportional to the conductivity  $\sigma(\Omega)$, $Im.D[\Omega]\propto \Omega \sigma(\Omega)$, the Raman spectrum contains the information about the conductivity.
For light  the momentum $q$  is negligible  and we can replace $W_{i}^{(p,a)}(\vec{k}+\vec{q},-\vec{k})\approx  U_{i}(\vec{k})$, $W_{i}^{(a,p)}(-\vec{k},\vec{k}+\vec{q})\approx  U^{*}_{i}(\vec{k})$ which represents the electronic spinor projection in the $i$ direction.   
Measuring the scattering light in the $j$ direction   for an incoming  light in the $i$ direction will be proportional to the density-density correlation  ($i$ and $j$ correspond to the Cartesian coordinates on the surface of the $TI$, the  incoming light vector $\vec{k}$ is almost  perpendicular  to the surface):
\begin{equation}
D_{j,i}[\vec{q},\Omega]=\int_{0}^{\infty}\,dt e^{i\omega t}\int\frac{d^2k}{(2\pi)^2 }\int\,\frac{d^2p}{(2\pi)^2 } U_{j}(\vec{k}) U_{i}(\vec{k})
U^{*}_{j}(\vec{p}) U^{*}_{i}(\vec{p})\langle C^{\dagger}(\vec{k}+\vec{q},t)C(\vec{k},t)C^{\dagger}(\vec{p}+\vec{q},t)C(\vec{p},0)\rangle] . 
\label{eq}
\end{equation}
The chiral properties of the surface electrons are  probed with polarized light \cite{Conder}.
The circular  polarized light   $e^{\eta}$, $\eta=\pm$   is expressed in terms of the  linear polarization  $\epsilon^{1}(\vec{k})$, $\epsilon^{2}(\vec{k})$. The  chiral correlation function $D_{-,+}[\vec{q},\Omega]$ is given in terms of the linear  polarized light and spinors $U_{i}(\vec{k})$: 
\begin{eqnarray} 
&&D_{-,+}[\vec{q},\Omega]= (D_{x,x}[\vec{q},\Omega]-D_{y,y}[\vec{q},\Omega])+i( D_{x,y}[\vec{q},\Omega]-D_{y,x}[\vec{q},\Omega]) , \nonumber\\&&
U_{i}(\vec{k})=\delta_{i,x}\sin[\chi(k_{x},k_{y}]-\delta_{i,y}\cos[ \chi(k_{x},k_{y}]\nonumber\\&&
Im.\sum_{i=x,y}\sum_{j=x,y}D_{j,i}[\vec{q}\rightarrow 0,\Omega]\propto \Omega \sigma(\Omega) . \nonumber \\&& 
\end{eqnarray}
In Fig. 2 we  plot the function    $D[\vec{q}\rightarrow 0,\Omega]\propto \Omega \sigma(\Omega)$  using the conductivity  given in Fig. 2.  $k_{F}l=15$ (the lower graph), $k_{F}l=10$,  $k_{F}l=5$(upper graph). The line shape in Fig. 2  is in qualitative agreement with the Raman line shift reported in ref \cite{Conder}.
\begin{figure}
\begin{center}
\includegraphics[width=7.25 in ]{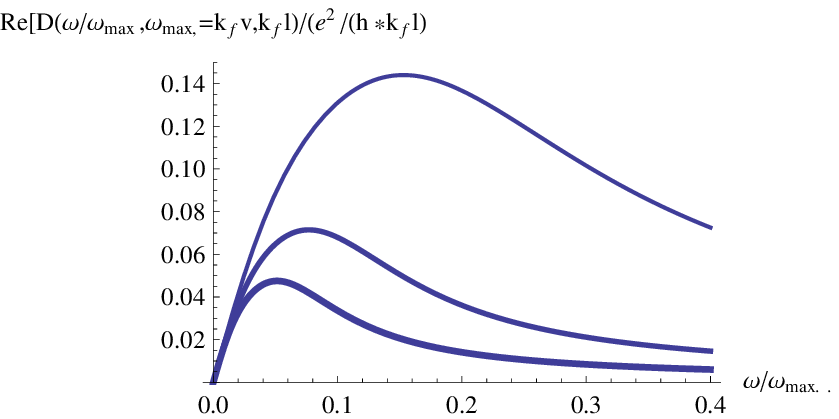}
\end{center}
\caption{The Raman shift   $D[\vec{q}\rightarrow 0,\Omega]\propto \Omega \sigma(\Omega)$ for   $k_{F}l=5$,  $k_{F}l=10$ , $k_{F}l=15$ } 
\end{figure}

\vspace{0.2 in}

\textbf{VIII Conclusions}

\vspace{0.2 in}

To conclude we have computed the conductivity for the conduction electrons on the surface of the $TI$ in the presence of a large chemical potential.  Due to the $TI$ topology the conduction band contains a half vortex at $\vec{k}=0$. As a result  the return probabilty vanishes and the quantum corrections are given by anti-localization effect as predicted for the symplectic ensemble. Using this theory we discuss the implications to the Raman scattering recently observed in $Bi_{2}Se_{3}$.


\begin{thebibliography}{99}

\bibitem{Volkov} B.A, Volkov  and O.A. Pankratov,  JETP Lett. \textbf{61}, 2015 (1988). 

\bibitem{Kane} C.L. Kane and E.J. Mele, Phys. Rev. Lett.\textbf{75}, 146802 (2005). 

\bibitem{Zhang} Xiao-Liang Qi and Shou-Cheng-Zhang, Rev.  Modern Physics \textbf{83}, 1057 (2011). 

\bibitem{David} D. Schmeltzer Phys.Rev.B \textbf{73}, 165301 (2006); Advances in Condensed Matter and Material Research, Editors Hans Geelvinck and Sjaak Reyst , volume \textbf{10}, chapter $9$,  pages $379-403$ (2011).

\bibitem{Culcer} D. Culcer cond-mat/1108.3076. 

\bibitem{Balatsky} R. Biswas and A.V. Balatsky,. Phys. Rev. B \textbf{81}, 23405 (2010). 

\bibitem{Hikami}  Shinobu Hikami, Phys. Rev. B\textbf{24}, 2671 (1981). 

\bibitem{Ando} Hidekatsu Suzura and Tsuneya Ando, Phys.Rev.Let. \textbf{108},076804 (2002) and  (preprint 2011). 

\bibitem{Hankiewicz} G. Thackhov and E.M. Hankiewicz,  cond-mat $01102.4512$  v4.

\bibitem{Stern} Zohar Ringel, Yaacov E. Kraus, and Ady Stern, Phys. Rev. B.\textbf{86}, 045102 (2012). 

\bibitem{Shen} Hai-Zhou Lu, Junren Shi, and Shun-Qing Shen, cond-mat$01101.5437$  v3.

\bibitem{Garate} Ion Garate and Leonid Glazman, Phys. Rev. B \textbf{86}, 035422 (2012). 
 
\bibitem{Raghu} S. Raghu, S.B. Chung, X.L. Qi, and S.-C. Zhang, Phys. Rev. Lett. \textbf{104},116401 (2010). 

\bibitem{Maggiore} Michelle Maggiore, {\it A modern Introduction to Quantum Field Theory},  (Oxford Univrsity Press, 2005). 

\bibitem{Abrikosov} A.A. Abrikosov, L.P. Gorkov, and I.E. Dzyaloshinski, {\it Methods of Quantum Field Theory In Statistical Physics}, `page 63. (Dover publications, 19--). 

\bibitem{Rammer} J. Rammer, {\it Quantum Field Theory of Non-Equilibrium States}, pages 377-389, (Cambridge University Press 2007).

\bibitem{Bergmann} G.Bergmann Phys.Rev.B \textbf{28},2914 (1983)

\bibitem{kaveh} D.Schmeltzer and M.Kaveh J.Phys.C \textbf{20},L175 (1987);Phys.Rev.B \textbf{37},9057 (1987)


\bibitem{Conder} V. Gnezdilov, Yu. G. Pashkevich, H. Berger, E. Pomjakushina, K. Conder, and P.Lemmens,  cond-mat $01108.2047$. 

\bibitem{Doniach} S. Donaich and E.H. Sondheimer, {\it Green's Functions for Solid State Physicists}, (Imperial College Press,  1988). 





\end{thebibliography}
\end{document}